\documentclass[prb, twocolumn, superscriptaddress, aps, longbibliography,
floatfix, reprint]{revtex4-2}

\usepackage[utf8]{inputenc}
\usepackage[T1]{fontenc}
\usepackage[table]{xcolor}
\usepackage{graphicx}
\usepackage{amsmath}
\usepackage{amssymb}
\usepackage{afterpage}
\usepackage{amsfonts}
\usepackage{braket}
\usepackage{textcomp}
\usepackage{bm}
\usepackage{siunitx}
\usepackage{booktabs}
\usepackage{hyperref}
\usepackage{placeins}
\usepackage{lipsum}
\usepackage{wasysym}
\usepackage{orcidlink}
\newcommand{\orcid}[1]{\orcidlink{#1}}
\hypersetup{
    colorlinks,
    linkcolor={blue!90!black},
    citecolor={blue!90!black},
    urlcolor={blue!90!black}
}
\DeclareUnicodeCharacter{0308}{\"{}}

\usepackage{newtxtext} 
\usepackage[subscriptcorrection,nosymbolsc,smallerops,bigdelims]{newtxmath} 
\DeclareMathAlphabet{\mathcal}{OMS}{cmsy}{m}{n} 
\DeclareMathAlphabet{\mathbcal}{OMS}{cmsy}{b}{n} 
\definecolor{colorSSSbase}{HTML}{1F77B4}
\definecolor{colorSSAbase}{HTML}{3B8CC4}

\definecolor{colorSASbase}{HTML}{D62728}
\definecolor{colorSAAbase}{HTML}{E04A4B}

\definecolor{colorASSbase}{HTML}{2CA02C}
\definecolor{colorASAbase}{HTML}{4DB44D}

\definecolor{colorAASbase}{HTML}{9467BD}
\definecolor{colorAAAbase}{HTML}{A982CC}

\colorlet{colorSSS}{colorSSSbase!80!white}
\colorlet{colorSSA}{colorSSAbase!80!white}

\colorlet{colorSAS}{colorSASbase!80!white}
\colorlet{colorSAA}{colorSAAbase!80!white}

\colorlet{colorASS}{colorASSbase!80!white}
\colorlet{colorASA}{colorASAbase!80!white}

\colorlet{colorAAS}{colorAASbase!80!white}
\colorlet{colorAAA}{colorAAAbase!80!white}
\begin{document}

\title{Cavity--waveguide coupling in phononic crystals}

\author{Jakub Rosi\'n{}ski\orcid{0000-0001-7941-0296}}
\email{jakub.rosinski@pwr.edu.pl}
\affiliation{Institute of Theoretical Physics, Wroc\l{}aw University of Science and Technology, Wroc\l{}aw, 50-370, Poland}

\author{Benjamin Mayer\orcid{0009-0005-0734-7909}}
\affiliation{Institute of Physics, University of M\"unster, 48149, M\"unster, Germany}

\author{Hubert J. Krenner\orcid{0000-0002-0696-456X}}
\affiliation{Institute of Physics, University of M\"unster, 48149, M\"unster, Germany}

\author{Pawe\l{} Machnikowski\orcid{0000-0003-0349-1725}}
\affiliation{Institute of Theoretical Physics, Wroc\l{}aw University of Science and Technology, Wroc\l{}aw, 50-370, Poland}

\email{email}

\begin{abstract}
Phononic crystal platforms provide a promising route toward scalable on-chip quantum networks, where mechanical excitations mediate interactions between solid-state qubits. A key building block of such architectures is the cavity--waveguide system, in which localized mechanical modes couple to propagating phononic modes. However, a quantitative understanding of the mechanisms governing this coupling remains incomplete. In this work, we investigate the interaction between localized modes of snowflake-type phononic crystal cavities and a phononic crystal waveguide using finite-element simulations and experimental measurements. We show that the coupling strength is primarily governed by the spatial overlap between the displacement fields of the corresponding isolated cavity and waveguide modes. After accounting for the effective-mass dependence of the normalized cavity displacement amplitude, we establish a strong correlation between the spatial overlap and the interaction strength across a wide range of mode combinations. Deviations from this leading-order behavior are associated with Bloch-phase effects, variations in waveguide group velocity, and intrinsic hybridization of waveguide modes. Furthermore, we provide experimental evidence for cavity--waveguide coupling in a GaAs phononic crystal membrane by observing a clear resonance in the spectral broadening of the photoluminescence emission from an embedded quantum dot, which serves as a local probe of the mechanical field. The observed resonance at approximately 398~MHz is in good agreement with the corresponding cavity resonance near 395~MHz predicted by finite-element simulations. Our results identify spatial mode overlap as a leading-order design parameter for cavity--waveguide coupling and provide practical guidelines for controlling interactions between localized and propagating modes in phononic crystal structures.
\end{abstract}

\date{\today}

\maketitle

\section{Introduction}
While photons are ideal for long-distance quantum communication, phonons offer distinct advantages for on-chip quantum communication~\cite{habraken2012continuous, lemonde2018phonon, krenner20262026}, making them a promising platform for building quantum computers based on robust solid-state spin qubits. Owing to their low phase velocity, phonons exhibit wavelengths that are orders of magnitude shorter than those of electromagnetic waves at the same frequency, enabling strong spatial confinement and enhanced interaction strengths in compact resonators and waveguides. Moreover, mechanical waves do not suffer from radiation losses into vacuum, as phonons do not propagate in vacuum.

Phononic waves provide an exceptionally versatile platform for coherent coupling to a wide range of solid-state quantum systems. Over the past decade, coherent phonon-mediated interactions have been demonstrated with semiconductor quantum dots (QDs)~\cite{metcalfe2010resolved,yeo2014strain,schulein2015fourier,weiss2018interfacing}, integrated photonic circuits~\cite{de2006compact, fuhrmann2011dynamic, kapfinger2015dynamic, weiss2016surface, balram2016coherent, kittlaus2021electrically, buhler2022chip, chen2023optomechanical}, superconducting circuits~\cite{o2010quantum,pirkkalainen2013hybrid,gustafsson2014propagating,chu2017quantum,bienfait2019phonon}, nitrogen-vacancy centers in diamond~\cite{arcizet2011single,kolkowitz2012coherent,macquarrie2013mechanical,ovartchaiyapong2014dynamic,teissier2014strain,macquarrie2015continuous,barfuss2015strong,macquarrie2015coherent,meesala2016enhanced,golter2016optomechanical,lee2016strain,golter2016coupling,lee2017topical,chen2018orbital}, and spin defects in silicon carbide~\cite{whiteley2019spin}. This broad compatibility highlights the potential of phonons as universal mediators of interactions in hybrid quantum architectures. Independent of the specific quantum platform, controlled routing and confinement of mechanical excitations constitute essential ingredients for scalable phononic networks~\cite{rabl2010quantum,lemonde2018phonon,kuzyk2018scaling}.

To build scalable quantum processors, large mechanical networks are required. Several physical platforms have been proposed to realize such networks. One approach uses one-dimensional chains of trapped ions, where Coulomb-coupled ions share collective vibrational modes that act as phonons~\cite{leibfried2003quantum}, mediating interactions between qubits. Another realization involves chains of mechanically coupled solid-state resonators~\cite{habraken2012continuous}, in which vibrational excitations propagate through direct physical connections. 

Both approaches, however, face fundamental scalability problems~\cite{monroe2013scaling}. As the number of elements increases, the single-phonon coupling strength decreases because the effective mass of the collective mechanical system grows—the coupling rate scales as \( g \sim 1/\sqrt{m} \), where \( m \) is the mass of the mechanical system. More critically, nearest-neighbor coupling leads to the formation of spectrally dense mechanical modes. The resulting crosstalk between overlapping modes prevents selective quantum control of individual vibrational states, severely limiting the size of a single mechanically connected module.

In optical quantum networks, high-fidelity state transfer between neighboring nodes can be achieved in cascaded~\cite{cirac1997quantum} architectures using chiral optical interactions~\cite{lodahl2017chiral}. An analogous approach for phononic networks would require chiral acoustic processes to enable unidirectional quantum state transfer. However, the lack of easily accessible chiral acoustic processes makes it difficult to implement such cascaded phononic quantum networks~\cite{habraken2012continuous,lemonde2018phonon}. This limitation motivates the search for alternative architectures that do not rely on unidirectional coupling.

To overcome these obstacles, architectures that divide a large mechanical network into small, closed mechanical subsystems have been proposed~\cite{kuzyk2018scaling,li2019honeycomblike}. In this implementation, localized mechanical resonators are connected via one-dimensional~\cite{kuzyk2018scaling} or two-dimensional~\cite{li2019honeycomblike} phononic crystal (PhC) waveguides, with band gaps engineered to ensure that phononic excitations are exchanged only between neighboring resonators. Within each resonator, solid-state spin qubits, such as color centers in diamond, couple to the local mechanical mode through sideband (phonon-assisted) transitions driven by external optical or microwave fields~\cite{golter2016optomechanical}, allowing phononic excitations to mediate interactions between qubits. Studying cavity–waveguide coupling is therefore essential for scalable phononic quantum networks.

A central building block of phononic networks is the cavity–waveguide system, in which a localized mechanical mode interacts with propagating modes of a PhC waveguide. Understanding and controlling this interaction is essential for the efficient transfer of phononic excitations between network nodes.

In this work, we investigate the coupling between localized mechanical modes of snowflake-type PhC cavities and a W1m PhC waveguide~\cite{rosinski2026couplingquantumdotselastic}. We show that the coupling strength is primarily governed by the spatial overlap of the uncoupled modes, with systematic deviations associated with waveguide dispersion, phase matching, and intrinsic hybridization effects. Using finite-element method (FEM), we extract the interaction strength from avoided crossings in the dispersion relation and correlate the resulting splittings with (i) spatial overlap integrals of the uncoupled modes and (ii) the group velocity of the relevant waveguide branches, thereby assessing the relative roles of structural and dispersive effects in the coupling. Furthermore, we provide experimental evidence of waveguide--cavity coupling in a GaAs PhC membrane, probed via an embedded QD that serves as a local sensor of the mechanical field through both deformation-potential and piezoelectric coupling~\cite{weiss2018interfacing,rosinski2026couplingquantumdotselastic}. Understanding the relative roles of these mechanisms in the coupling strength provides guidance for targeted optimization of phononic structures toward efficient and scalable quantum networks.

The paper is structured as follows. In Sec.~\ref{sec:model} we introduce the system geometry and the theoretical framework used to quantify the coupling. In Sec.~\ref{sec:results} we present the numerical results and establish the scaling of the coupling strength with spatial overlap, followed by an analysis of deviations arising from phase matching, dispersion, and waveguide hybridization. In Sec.~\ref{sec:results_experiment} we provide experimental evidence supporting the cavity--waveguide coupling mechanism. We conclude the paper in Sec.~\ref{sec:conclusion}. The Appendix provides additional information.

\section{System, Model, and Methods}\label{sec:model}
In this section, we describe the design and modeling of the isolated PhC cavities and waveguide, as well as the coupled cavity--waveguide system. We first present the geometry and symmetry properties of the isolated snowflake-based PhC cavities (L0, L1, L2) and of the W1m waveguide (Sec.~\ref{sec:system_isolated}). We then describe the coupled cavity–waveguide structures used in the eigenfrequency simulations and the procedure for extracting the coupling strength from avoided crossings (Sec.~\ref{sec:system_coupled}). Finally, we introduce two key quantities that allow us to assess structural and dispersive contributions to the coupling: the spatial overlap integral $\eta$ between the uncoupled displacement fields and the group velocity $v_{\rm g}$ of the waveguide mode (Sec.~\ref{sec:system_coupling}).

\subsection{Isolated cavities and waveguide}\label{sec:system_isolated}

The PhC structure considered here consists of a hexagonal array of snowflake-shaped inclusions in a thin (001)-oriented GaAs plate, adapted from our previous work~\cite{rosinski2026couplingquantumdotselastic}, designed to exhibit a phononic band gap centered around 3~GHz. The selected material platform meets the integration requirements for QDs commonly used in optomechanical systems, such as InAs or AlGaAs. 

\begin{figure*}[tb]
	\includegraphics[width=\textwidth]{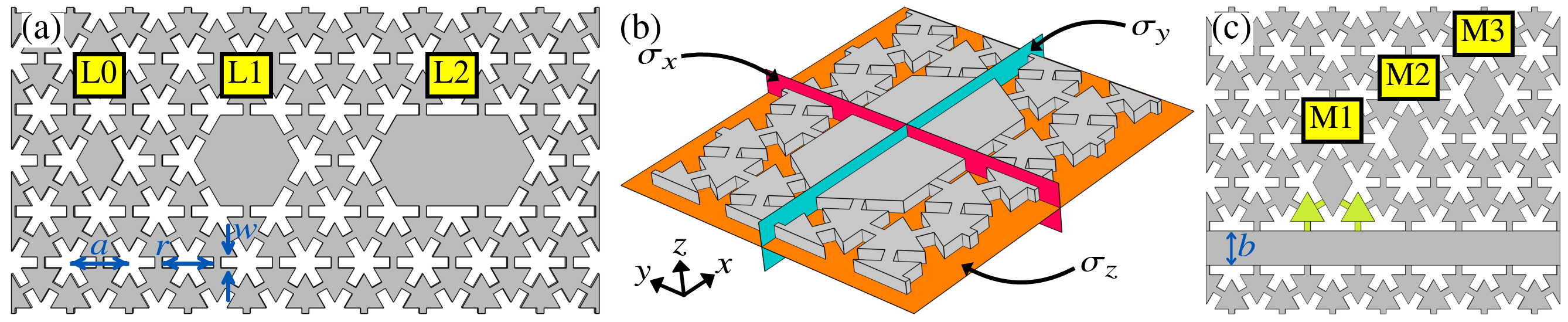}
	\caption{(a) Three phononic cavity designs (L0, L1, L2) realized in the snowflake-based PhC, with the geometric parameters $a$, $r$, and $w$ indicated. (b) Symmetry planes of the cavities. (c) Schematic of the L0 cavity coupled to the W1m waveguide. The waveguide core width is controlled by the parameter $b$, while M1, M2, and M3 indicate three cavity positions corresponding to different cavity--waveguide separations. The yellow regions indicate the integration domains used to evaluate the overlap integral between the L0 cavity and the W1m waveguide, illustrated for the M1 separation.}
	\label{fig:fig1}
\end{figure*}

The geometry is characterized by the lattice constant $a$, the snowflake radius $r$, the arm width $w$, and the membrane thickness $h$. The in-plane geometric parameters $a$, $r$, and $w$ are indicated in Fig.~\ref{fig:fig1}(a), while \(h\) denotes the membrane thickness. The parameter values are the same as those used in Ref.~\cite{rosinski2026couplingquantumdotselastic}.

Three distinct phononic cavity designs, labeled L0, L1, and L2, are investigated in this study. They are created within the same snowflake-based PhC by locally modifying the lattice in the following way: the L0 cavity is formed by removing two opposite arms of a central snowflake inclusion, the L1 cavity by removing one complete snowflake inclusion, and the L2 cavity by removing two adjacent snowflake inclusions. These modifications are illustrated in Fig.~\ref{fig:fig1}(a).

The phononic waveguide used in this work is a W1m-type line-defect waveguide, originally proposed in Ref.~\cite{rosinski2026couplingquantumdotselastic}. Its core width is given by $b=\sqrt{3}a-2\Delta-w$, with the lateral offset fixed at $\Delta=\sqrt{3}a/4$, as illustrated in Fig.~\ref{fig:fig1}(c). 

Each cavity possesses three mirror planes: a horizontal mid-plane $\sigma_z$ at $z=0$ and two vertical planes, $\sigma_y$ (at $y=0$) and $\sigma_x$ (at $x=0$), as illustrated in Fig.~\ref{fig:fig1}(b). Cavity eigenmodes are classified by their parity under reflections with respect to these planes. The mode symmetry is denoted by the triplet $(p_z,p_y,p_x)$, where each entry is S (symmetric/even) or A (antisymmetric/odd). Because the waveguide lacks the $\sigma_x$ mirror symmetry, its modes are classified only with respect to the $\sigma_z$ and $\sigma_y$ planes.

For the isolated cavities, eigenfrequency calculations are performed using a supercell with in-plane dimensions of $11a$ along $x$ and $10a\sqrt{3}/2$ along $y$, with standard periodic boundary conditions applied in both directions. For the isolated waveguide, calculations are performed using a unit cell with dimensions of $a$ along the propagation direction $x$ and $10a\sqrt{3}/2$ along $y$, with Bloch--Floquet periodic boundary conditions characterized by the wavenumber $k_x$ along $x$ and standard periodic boundary conditions along $y$. In all cases, traction-free mechanical boundary conditions are imposed on the membrane surfaces at $z=\pm h/2$.

\subsection{Coupled cavity–waveguide structures}\label{sec:system_coupled}
To study the interaction between the cavities and the waveguide, we consider systems in which one of the cavities (L0, L1 or L2) is placed in close proximity to the W1m waveguide. The coupling strength is controlled by varying the number of lattice periods separating the waveguide from the cavity. In this work, we investigate configurations with M1, M2 and M3. The corresponding geometries are illustrated in Fig.~\ref{fig:fig1}(c). In these coupled systems, the asymmetric placement of the cavity next to the W1m waveguide reduces the overall symmetry of the structure to a single mirror plane --- the horizontal mid-plane $\sigma_z$ at $z = 0$. As a result, the interaction between the cavity and the waveguide is governed by symmetry-based selection rules: a given cavity mode couples only to waveguide modes that have the same parity with respect to $\sigma_z$.

For the L0 cavity, eigenfrequency calculations are performed for all three separation configurations (M1--M3) using a large supercell containing both the waveguide segment and the cavity. Selected configurations are additionally considered for the L1 cavity. The supercell has the same in-plane dimensions as used for the isolated cavities: width $11a$ in the $x$-direction and height $10a\sqrt{3}/2$ in the $y$-direction. Periodic boundary conditions are applied along the transverse direction $y$, while Bloch--Floquet periodic boundary conditions with a specified wavenumber component $k_x$ are imposed along the waveguide propagation direction $x$. Because the coupled calculations employ a supercell of length $11a$ along the waveguide direction, the waveguide Bloch bands are folded into the reduced Brillouin zone (BZ). The relation between the folded wavevector and the corresponding wavevector of the primitive waveguide unit cell is described in Appendix~\ref{sec:appendix_folding}. Traction-free mechanical boundary conditions are applied to the membrane surfaces at $z=\pm h/2$.

\subsection{Quantification of the coupling strength}\label{sec:system_coupling}
This section establishes the central framework used throughout the paper to interpret the cavity--waveguide interaction. 
Due to Floquet boundary conditions, the cavity modes are represented as flat-band (constant frequency) modes, reflecting their bound character.
Whenever the frequency of a cavity mode lies close to that of a symmetry-allowed waveguide branch, the two modes hybridize. This hybridization manifests as an avoided crossing (anticrossing) in the system's dispersion diagram. The interaction strength $g$ between the cavity mode and the waveguide mode is then extracted directly from the frequency splitting at this anticrossing point
\begin{equation}
    2g = f_+ - f_-,
    \label{eq:couplingStrength}
\end{equation}
where $f_+$ and $f_-$ are the frequencies of the upper and lower hybrid modes, respectively.

The coupling strength $g$ extracted from the frequency splitting in the coupled eigenfrequency calculations [see Eq.~\eqref{eq:couplingStrength}] provides a direct measure of the interaction, but does not immediately reveal its microscopic origin. To gain deeper physical insight into the origin of the interaction, we primarily investigate the spatial overlap integral $\eta$ between the displacement fields of the uncoupled cavity and waveguide modes and subsequently examine additional mechanisms responsible for deviations from the overlap-based scaling.

The parameter $\eta$ is defined as the absolute value of the spatial overlap integral between the normalized displacement fields of the isolated cavity mode and the isolated waveguide mode
\begin{equation}
\eta = \left| \int_{V_{\rm OR}} \tilde{\mathbf{u}}_{\rm cav}(\mathbf{r}) \cdot \tilde{\mathbf{u}}_{\rm wg}^*(\mathbf{r}) \, dV \right|,
\label{eq:overlap1}
\end{equation}
where $V_{\rm OR}$ is the overlap region between the L0 cavity and the W1m waveguide [illustrated in Fig.~\ref{fig:fig1}(c)], $\tilde{\mathbf{u}}_{\rm cav}$ is the normalized cavity mode displacement, and $\tilde{\mathbf{u}}_{\rm wg}$ is the normalized mode of the waveguide (per unit cell). Although the overlap integral $\eta$ is formally defined for any cavity, in the following analysis we focus on the smallest L0 cavity because it exhibits the strongest confinement and the largest number of well-resolved anticrossings with the waveguide branches. The same procedure can be applied to L1 and L2.

The cavity and waveguide displacement fields are normalized to have the same total time-averaged mechanical energy of the mode, defined as
\begin{equation}
\hbar\omega_0 = \frac{1}{2} \int_V \rho(\mathbf{r}) \omega^2 |\mathbf{u}(\mathbf{r})|^2 \, dV = \dfrac{1}{2} m_{\text{eff}} \omega^2 \alpha^2,
\label{eq}
\end{equation}
where $\rho$ is the mass density, $\omega$ the angular eigenfrequency, \( \alpha = \max(|\mathbf{u}(\mathbf{r})|) \), and $\omega_0=2\pi\times3.14~\mathrm{GHz}$ is chosen as the reference angular frequency. The effective mass for the mode is defined as~\cite{eichenfield2009modeling,aspelmeyer2013cavity}
\begin{equation}
    m_{\text{eff}} = \dfrac{ \int \rho |u(r)|^2 dV }{\text{max}[ |u(r)| ^2]}.
    \label{eq:effectiveMass}
\end{equation}
For the cavity mode the integral is taken over the supercell. For the waveguide mode the energy is computed per unit cell, i.e.\ the total energy in the supercell is divided by $N = 11$ before normalization. The normalized displacement fields are written in component form as
\begin{align}
\tilde{\mathbf{u}}_{\rm cav}(\mathbf{r}) &= \begin{pmatrix} 
|\tilde u_x^{\rm cav}| e^{i \phi_x^{\rm cav}} \\ 
|\tilde u_y^{\rm cav}| e^{i \phi_y^{\rm cav}} \\ 
|\tilde u_z^{\rm cav}| e^{i \phi_z^{\rm cav}} 
\end{pmatrix} e^{i \phi_{\rm cav, global}}, \notag \\
\tilde{\mathbf{u}}_{\rm wg}(\mathbf{r}) &= \begin{pmatrix} 
|\tilde u_x^{\rm wg}| e^{i \phi_x^{\rm wg}} \\ 
|\tilde u_y^{\rm wg}| e^{i \phi_y^{\rm wg}} \\ 
|\tilde u_z^{\rm wg}| e^{i \phi_z^{\rm wg}} 
\end{pmatrix} e^{i \phi_{\rm wg, global}} \, e^{-i k_x x}.
\end{align}

Substituting these expressions into the overlap integral yields
\begin{equation}
\eta = 
\left| \sum_{j=x,y,z} \int_{V_{\rm OR}} 
|\tilde u_j^{\rm cav}| \, |\tilde u_j^{\rm wg}| \,
e^{i \left[ \Delta \phi_j(\mathbf{r}) + \Delta \phi_{\rm global} \right]} \, dV \right|,
\label{eq:overlap2}
\end{equation}
where
\[
\begin{aligned}
\Delta \phi_j(\mathbf{r}) 
    &= \phi_j^{\rm cav}(\mathbf{r}) - \phi_j^{\rm wg}(\mathbf{r}) + k_x x, \\
\Delta \phi_{\rm global} 
    &= \phi_{\rm cav, global} - \phi_{\rm wg, global}.
\end{aligned}
\]
This decomposition allows separating local amplitude overlap from phase matching for each Cartesian component, while global phase offsets can be factored out.

While $\eta$ quantifies the structural and phase overlap, the dispersive properties of the waveguide may provide an additional contribution to the cavity--waveguide interaction. In a one dimensional waveguide, the density of states (DOS) is inversely proportional to the group velocity
\begin{equation}
\mathrm{DOS}(\omega_c) \propto \frac{1}{|v_{\rm g}(\omega_c)|},
\qquad
v_{\rm g}(\omega_c)
=
\left.
\frac{d\omega}{dk_x}
\right|_{\omega_c}.
\label{eq:DOS}
\end{equation}
Reduced group velocity in periodic waveguides is associated with the slow-wave regime, particularly near band edges where strong dispersion occurs~\cite{hatanaka2014phonon, modica2020slow}. The resulting slow propagation can promote spatial concentration of the propagating acoustic energy~\cite{sun2010resonant}. This slow-phonon regime is analogous to the well-established slow-light phenomenon in photonic crystal waveguides~\cite{baba2008slow}. Motivated by these slow-wave effects, we expect reduced group velocity to favor stronger cavity--waveguide interaction in our system. To examine this contribution, we extract the group velocity $v_{\rm g}$ from the waveguide dispersion at the cavity resonance frequencies.

\section{Modeling results} \label{sec:results}
We analyze the cavity--waveguide coupling in a sequence of steps aimed at identifying the dominant physical mechanisms governing the interaction. We first characterize the isolated cavity and waveguide modes. We then extract coupling strengths from avoided crossings in the coupled dispersion and demonstrate that they are primarily determined by the spatial overlap of the uncoupled modes. Finally, we analyze systematic deviations from this scaling and attribute them to phase matching, waveguide dispersion, and intrinsic hybridization effects.

\subsection{Isolated structures}\label{sec:results_isolated}

\begin{figure*}[tb]
	\includegraphics[width=\textwidth]{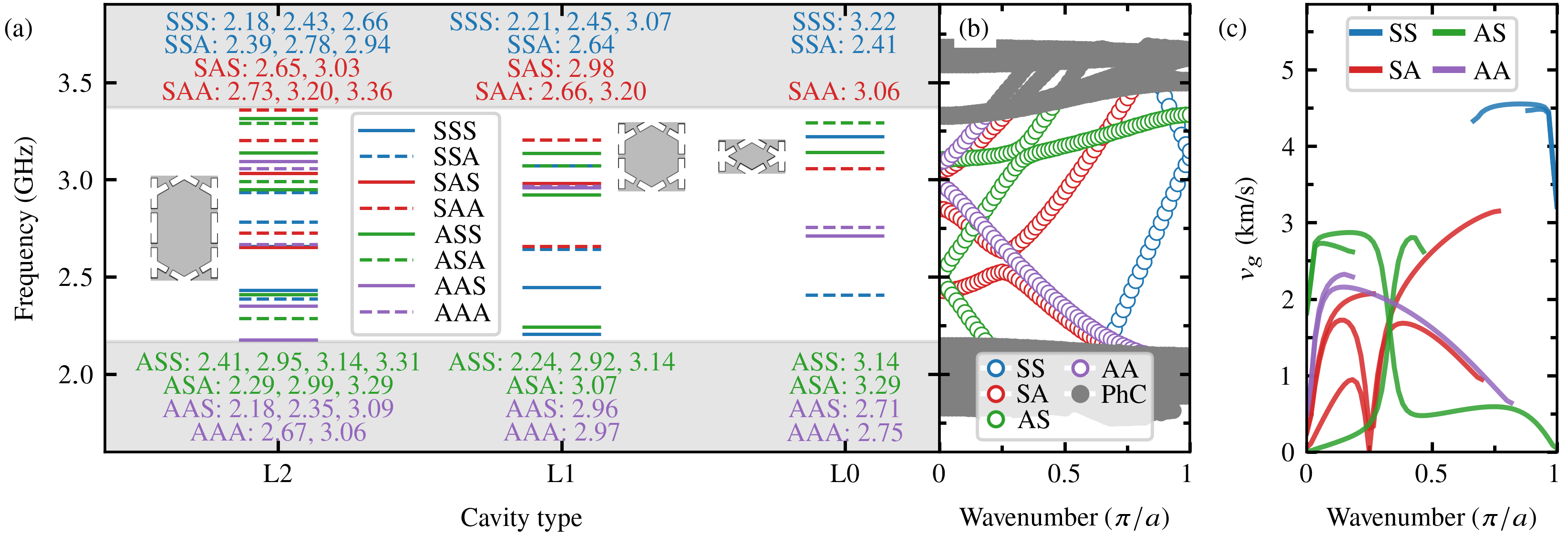}
	\caption{Cavity and waveguide modes in the PhC. (a) Eigenfrequencies of the localized cavity modes for the three cavity designs L2, L1, and L0. Insets show the corresponding cavity geometries. (b) Dispersion relations of the guided modes in the PhC waveguide. The gray shaded regions in (a) and (b) denote PhC bands. (c) Group velocities $v_{\rm g}$ of the four waveguide modes versus wavenumber. Colors indicate the mode symmetry with respect to the $\sigma_z$ and $\sigma_y$ mirror planes. For the cavity modes in panel (a), which additionally possess $\sigma_x$ symmetry, solid and dashed lines distinguish modes that are symmetric (S) and antisymmetric (A) with respect to $\sigma_x$, respectively.}
	\label{fig:fig2}
\end{figure*}
Figure~\ref{fig:fig2} summarizes the modal properties of the isolated cavity and waveguide structures forming the basis for the coupling analysis presented in the next section. All results were obtained from FEM eigenfrequency simulations performed for the cavity and waveguide separately.

The eigenfrequencies of localized cavity modes are shown in Fig.~\ref{fig:fig2}(a) for three progressively reduced defect cavities (L2, L1, and L0). The largest cavity (L2) supports 23 localized modes, while the L1 and L0 designs support 14 and 7 modes, respectively. The systematic reduction in the number of bound states with decreasing cavity size reflects the stronger spatial confinement provided by smaller defects. Consistent with this strong confinement, the L0 cavity modes exhibit low effective masses, ranging from approximately $17$ to $169$~fg, with corresponding zero-point motion amplitudes reaching approximately $13$~fm. The effective masses and zero-point motion amplitudes of all L0 cavity modes are summarized in Appendix~\ref{sec:appendix_cavityOpto}.

The dispersion relations of the guided modes in the PhC waveguide are presented in Fig.~\ref{fig:fig2}(b). Propagating modes belonging to all four symmetry classes (SS, SA, AS, and AA) are found within the considered frequency range, providing multiple spectral overlaps with the cavity resonances. The SS branch exhibits the steepest dispersion for large wavenumbers ($k$ approaching $\pi/a$), corresponding to the highest group velocity, which reaches approximately 4.5~km/s, as shown in Fig.~\ref{fig:fig2}(c). In contrast, the SA and AS branches display pronounced avoided crossings around $ka/\pi \approx 0.25$ and $0.33$, respectively, where modes of identical symmetry hybridize, leading to local flattening of the dispersion. As a consequence, the group velocity is strongly reduced in the vicinity of the anticrossings and approaches near-zero values for the SA branch, indicating the possibility of slow-sound behavior in the waveguide. Apart from these extrema, the group velocities of most guided modes lie in the range of approximately 0.5–3~km/s.

\subsection{Coupled structures}\label{sec:results_coupled}
We now examine the interaction between the L0 cavity and the W1m waveguide. The L0 design supports seven localized mechanical modes inside the phononic band gap. Due to the high density of waveguide branches in this frequency range, each cavity mode spectrally overlaps with two or three waveguide branches. This results in multiple anticrossings appearing in the eigenfrequency spectrum of the coupled system. 
The branches with constant-energy asymptotics away from resonances correspond to localized cavity modes.

\begin{figure}[tb]
	\includegraphics[width=\columnwidth]{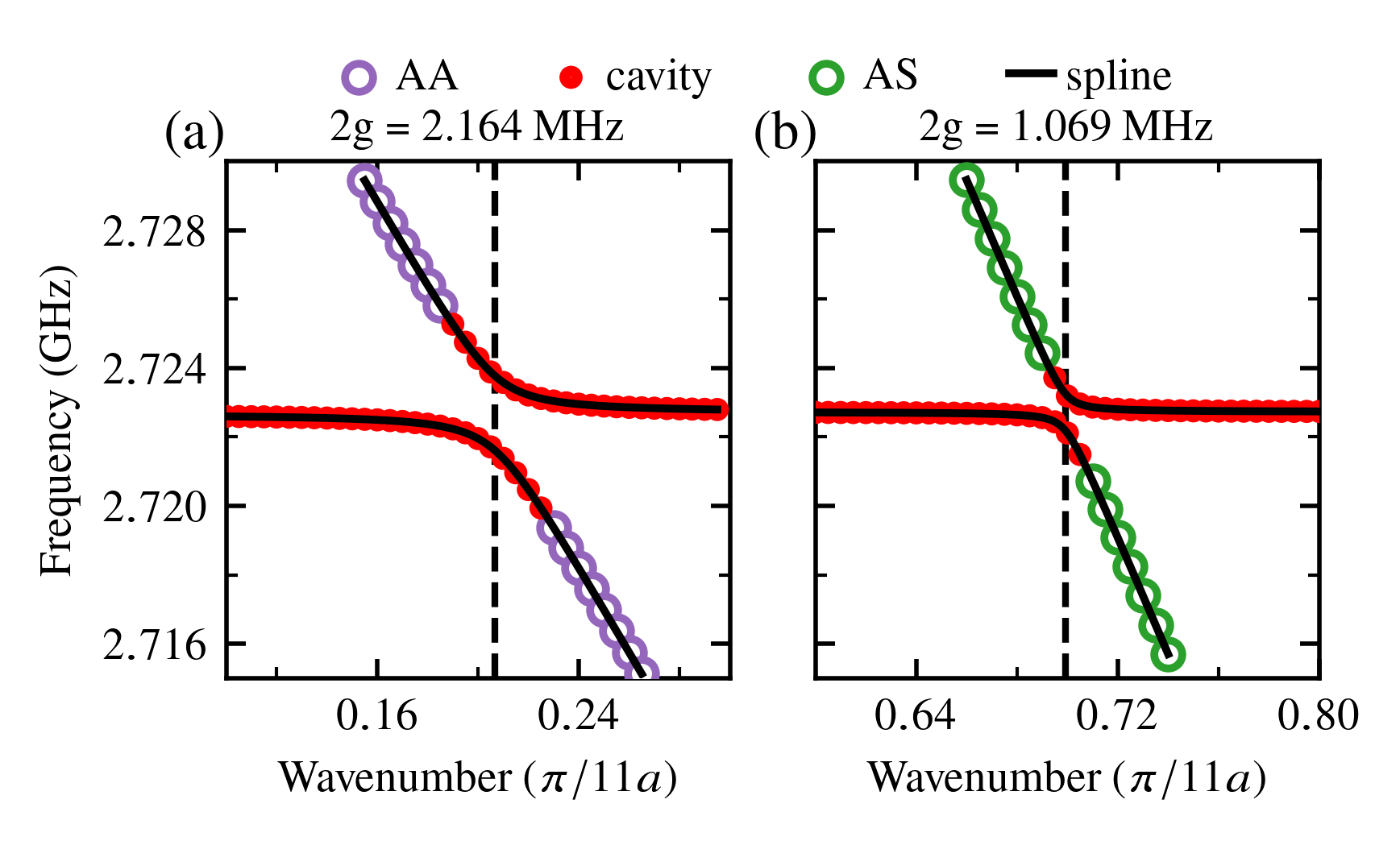}
	\caption{Dispersion diagram showing two anticrossings between the AAS cavity mode and waveguide modes with matching symmetry, corresponding to identical parity with respect to the $\sigma_z$ plane: the AA branch in panel~(a) and the AS branch in panel~(b). Red dots denote modes identified as cavity-like, green circles correspond to the AS waveguide mode, and purple circles to the AA waveguide mode. Black solid lines are cubic-spline fits used to determine the minimum frequency splitting. Blue dashed vertical lines mark the Bloch wavenumbers at which the anticrossings occur, and the extracted coupling strengths $2g$ are indicated above each panel.}
	\label{fig:fig3}
\end{figure}

Figure~\ref{fig:fig3} illustrates the procedure used to extract the coupling strength $2g$ from avoided crossings in the dispersion relation, shown here for a representative example involving the AAS cavity mode, with further details provided in Appendix~\ref{sec:appendix_coupling_extraction}. Two distinct anticrossings arise from the interaction of the AAS mode with the AA waveguide branch [Fig.~\ref{fig:fig3}(a)] and the AS waveguide branch [Fig.~\ref{fig:fig3}(b)]. Notably, the interaction of the AAS mode with the AA branch yields a significantly larger coupling strength ($2g = 2.164$~MHz) than with the AS branch ($2g = 1.069$~MHz). This difference appears to be primarily governed by the larger spatial overlap integral $\eta$ between the AAS cavity mode and the AA waveguide mode, while the group velocities and Bloch wavenumbers are comparable in this case. The ratio $2g_{\rm AA}/2g_{\rm AS}$ therefore closely follows the ratio $\eta_{\rm AA}/\eta_{\rm AS}$. A similar behavior is observed for the AAA mode, which interacts with the same branches. The complete set of extracted coupling strengths is summarized in Table~\ref{tab:final_antycrossings_reordered}. The values span two orders of magnitude, ranging from below 0.05~MHz to several MHz.

\begin{table}[tb]
\centering
\begin{tabular}{@{}ccccccc@{}}
\hline
Cav. & Wg. & $k$ & $\tilde{k}_L$ & $v_{\rm g}$ & $2g$ & $\eta$ \\
mode & mode & $(\pi/a)$ & $(\pi/(11a))$ & (m/s) & (MHz) & (\AA$^5$) \\
\hline

\rowcolor{colorSSA} 
SSA & SA & 0.354 & 0.11937 & 1658.86 & 5.363 & 3.136 \\
\rowcolor{colorSSA}
SSA & SS & 0.772 & 0.49542 & 4533.07 & 0.334 & 0.209 \\

\rowcolor{colorAAS}
AAS & AS & 0.118 & 0.69923 & 2857.87 & 1.067 & 1.058 \\
\rowcolor{colorAAS}
AAS & AA & 0.201 & 0.20672 & 2131.96 & 2.164 & 2.057 \\

\rowcolor{colorAAA}
AAA & AS & 0.144 & 0.41005 & 2866.58 & 1.792 & 0.717 \\
\rowcolor{colorAAA}
AAA & AA & 0.167 & 0.18045 & 2152.52 & 5.488 & 1.775 \\

\rowcolor{colorSAA}
SAA & SA & 0.063 & 0.68900 & 1424.94 & 3.527 & 0.357 \\
\rowcolor{colorSAA}
SAA & SA & 0.579 & 0.37368 & 2838.70 & 1.286 & 0.247 \\
\rowcolor{colorSAA}
SAA & SS & 0.986 & 0.83478 & 3679.47 & 0.167 & 0.049 \\

\rowcolor{colorASS}
ASS & AA & 0.062 & 0.68875 & 2012.03 & 0.247 & 0.795 \\
\rowcolor{colorASS}
ASS & AS & 0.293 & 0.76668 & 694.53 & 0.042 & 0.536 \\
\rowcolor{colorASS}
ASS & AS & 0.406 & 0.44788 & 527.35 & 0.111 & 0.282 \\

\rowcolor{colorSSS}
SSS & SA & 0.188 & 0.08313 & 2020.68 & 7.702 & 1.700 \\

\rowcolor{colorASA}
ASA & AA & 0.172 & 0.10186 & 2301.21 & 1.775 & 6.279 \\
\rowcolor{colorASA}
ASA & AS & 0.415 & 0.57052 & 2802.03 & 1.547 & 5.129 \\
\rowcolor{colorASA}
ASA & AS & 0.863 & 0.48858 & 512.06 & 0.947 & 2.171 \\

\hline
\end{tabular}
\caption{Summary of cavity--waveguide anticrossings. Listed are the cavity mode symmetry, dominant displacement-field component of the cavity, waveguide mode symmetry, Bloch wavenumber $k$ for a single unit cell, and the corresponding folded wavenumber $\tilde{k}_L$ for a supercell with $N=11$ (see Appendix~\ref{sec:appendix_folding} for the relation between $k$ and $\tilde{k}_L$). The table further reports the dominant waveguide displacement component, group velocity $v_{\rm g}$, coupling strength $2g$, overlap integral $\eta$ (\AA$^5$).}
\label{tab:final_antycrossings_reordered}
\end{table}

\begin{figure}[tb]
	\includegraphics[width=\columnwidth]{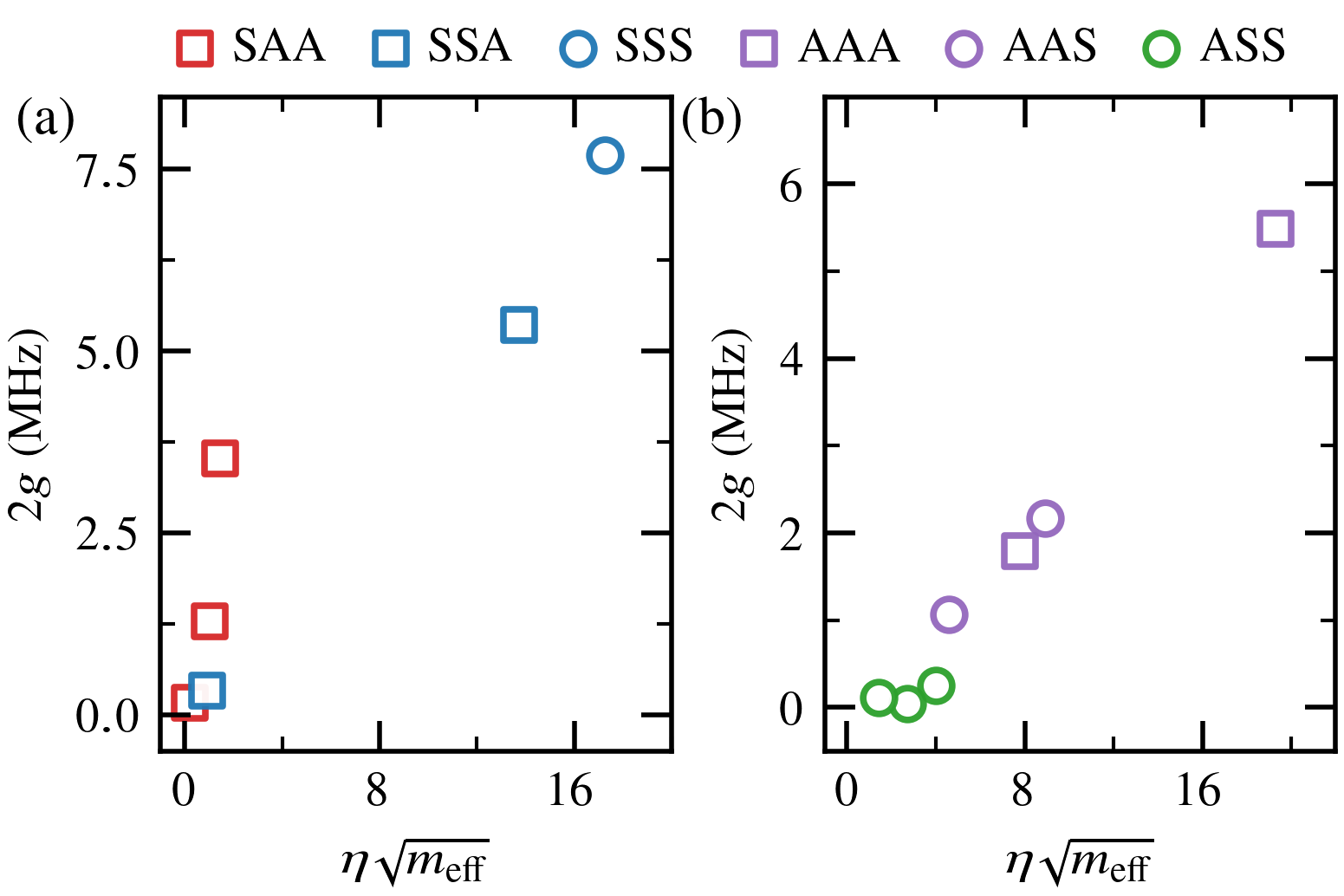}
	\caption{Correlation between the extracted coupling strength \(2g\) and the rescaled spatial overlap integral \(\eta\sqrt{m_{\rm eff}}\) for all anticrossings of the L0 cavity except those involving the ASA mode. Data are grouped by parity with respect to the horizontal mirror plane $\sigma_z$, with symmetric modes shown in panel~(a) and antisymmetric modes in panel~(b), corresponding to the first symmetry labels S and A, respectively. The corresponding Pearson correlation coefficients are $r=0.93$ and $r=0.99$, respectively. }
	\label{fig:fig4}
\end{figure}

A direct comparison between the overlap integral $\eta$ and the extracted coupling strengths $2g$ across different modes does not reveal a universal scaling. Since the displacement fields are normalized to the same energy, their amplitudes depend on the effective mass of the cavity mode, with the corresponding values summarized in Appendix~\ref{sec:appendix_cavityOpto}. Including the factor $\sqrt{m_{\rm eff}}$ removes the effective-mass dependence of the normalized displacement amplitude. Figure~\ref{fig:fig4} shows \(2g\) plotted against \(\eta\sqrt{m_{\rm eff}}\) for all anticrossings of the L0 cavity except those involving the ASA mode. The ASA mode lies at the edge of the phononic band gap and is therefore strongly delocalized [see Appendix~\ref{sec:appendix_fields}]; its spatial profile violates the assumption of a well-confined cavity mode underlying the overlap-integral analysis and the ASA mode was therefore excluded from this analysis. The remaining data are separated according to parity with respect to the horizontal mirror plane $\sigma_z$, with symmetric modes corresponding to the first symmetry label S and antisymmetric modes to A. The two classes are shown in Fig.~\ref{fig:fig4}(a) and (b), respectively, and exhibit strong linear correlations, with Pearson correlation coefficients of 0.93 and 0.99, respectively. These high values indicate that, once the effective-mass dependence is accounted for, the coupling strength $2g$ is strongly correlated with the spatial overlap integral $\eta$, suggesting that it constitutes the leading-order contribution. The remaining scatter may reflect secondary contributions from group velocity, phase-matching conditions, and waveguide hybridization, as discussed below.

\subsection{Deviations from overlap scaling}
While the overlap-based scaling captures the dominant trend, clear deviations arise in specific cases where additional physical factors become relevant. One such case is the AAA cavity mode interacting with the AA and AS waveguide branches. Although the Bloch wavevectors are very close ($k=0.144\,\pi/a$ for AS and $k=0.167\,\pi/a$ for AA), the ratio of the extracted coupling strengths, $2g(\mathrm{AA})/2g(\mathrm{AS}) = 3.06$, is noticeably larger than the corresponding overlap-integral ratio, $\eta(\mathrm{AA})/\eta(\mathrm{AS}) = 2.48$. Interestingly, including the group velocity through the heuristic factor $1/\sqrt{v_g}$ increases the corresponding ratio to $(\eta/\sqrt{v_g})_{\rm AA}/(\eta/\sqrt{v_g})_{\rm AS} = 2.86$, bringing it closer to the observed coupling ratio of $3.06$. This suggests that local waveguide dispersion may contribute to the remaining discrepancy.
 
A more complex situation, involving both Bloch-phase and group-velocity effects, is observed for the SAA cavity mode interacting with the SA waveguide branch at two markedly different wavevectors. Despite comparable values of the overlap integral $\eta$, the extracted coupling strengths differ significantly, with the strongest coupling for the SAA mode occurring at the smaller wavevector \(k = 0.063\,\pi/a\), where \(2g = 3.527\)~MHz. At this point, two favourable conditions coincide. First, the Bloch wavelength is much larger than the separation between the two successive islands ($\lambda_{\rm{B}} = 2\pi/k \approx 32a$), so that their Bloch phases remain nearly identical. This phase relationship is consistent with the even parity of the dominant $u_x$ component in the isolated SAA cavity mode (see Fig.~\ref{fig:fig11} in Appendix~\ref{sec:appendix_fields}) with respect to $x=0$, enabling constructive interference between the couplings to both islands. Second, the waveguide group velocity is reduced (\(v_g = 1425\) m/s versus \(2839\) m/s at \(k = 0.579\,\pi/a\)), which may provide an additional contribution to the observed difference in coupling strength. In contrast, at $k = 0.579\,\pi/a$ ($2g = 1.286$~MHz), the much shorter Bloch wavelength $\lambda_{\rm{B}} \approx 3.5a$ produces a rapid phase variation and a nearly $\pi$-phase shift between the two islands in adjacent periods. Although the resulting scalar overlap $\eta$ remains comparable, the two cases exhibit markedly different spatial phase patterns. This suggests that the displacement-overlap metric used here, while explicitly containing the Bloch phase, does not fully represent the phase-sensitive elastic interaction between the cavity and waveguide modes. The different Bloch-phase structure may therefore contribute to the significantly weaker coupling observed at the larger wavevector.

\begin{figure}[tb]
    \centering
    \includegraphics[width=\columnwidth]{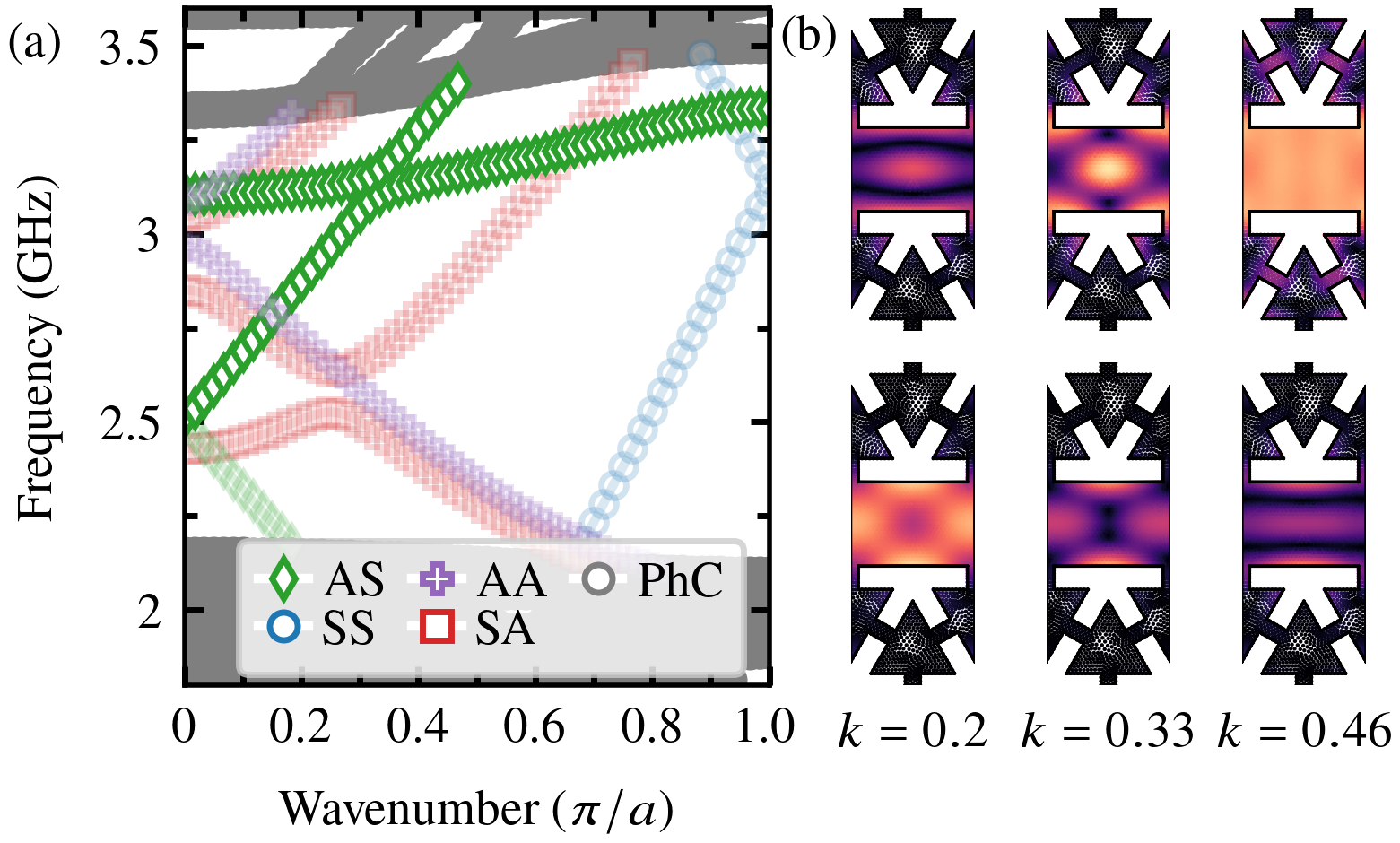}
    \caption{Hybridization of two interacting AS waveguide modes. (a) Waveguide dispersion, with the two relevant AS branches highlighted against the remaining modes. (b) Corresponding normalized total displacement amplitudes $|\mathbf{u}|$ for the upper and lower hybridizing branches at three representative wavevectors, $k=0.1$, $0.25$, and $0.4$. The wavevector $k$ is given in units of $\pi/a$. The evolution and exchange of the spatial field profiles across the avoided crossing illustrate the hybridization between the two waveguide modes.}
    \label{fig:fig5}
\end{figure}
Another source of deviation from the overlap-based scaling arises from hybridization within the waveguide modes themselves. For the ASS cavity mode, the interaction with the AA waveguide branch follows the expected overlap trend (larger $\eta$ gives larger $2g$). However, the two anticrossings with the AS branch show a much weaker correlation between $2g$ and $\eta$. Figure~\ref{fig:fig5} shows the evolution of the displacement fields of the two AS waveguide branches across their avoided crossing. As the wavevector is varied through the anticrossing, the two branches progressively exchange their spatial mode profiles, directly demonstrating the hybridization of the waveguide modes. This occurs because, at those wavevectors, the AS waveguide mode lies near an avoided crossing with another AS branch and undergoes strong hybridization. The resulting redistribution of the displacement field modifies the interaction of the waveguide mode with the localized cavity mode, even when the nominal overlap integral remains similar. Thus, intrinsic waveguide hybridization constitutes a distinct secondary mechanism that breaks the simple overlap scaling.

\subsection{Distance dependence}
\begin{figure}[tb]
	\includegraphics[width=\columnwidth]{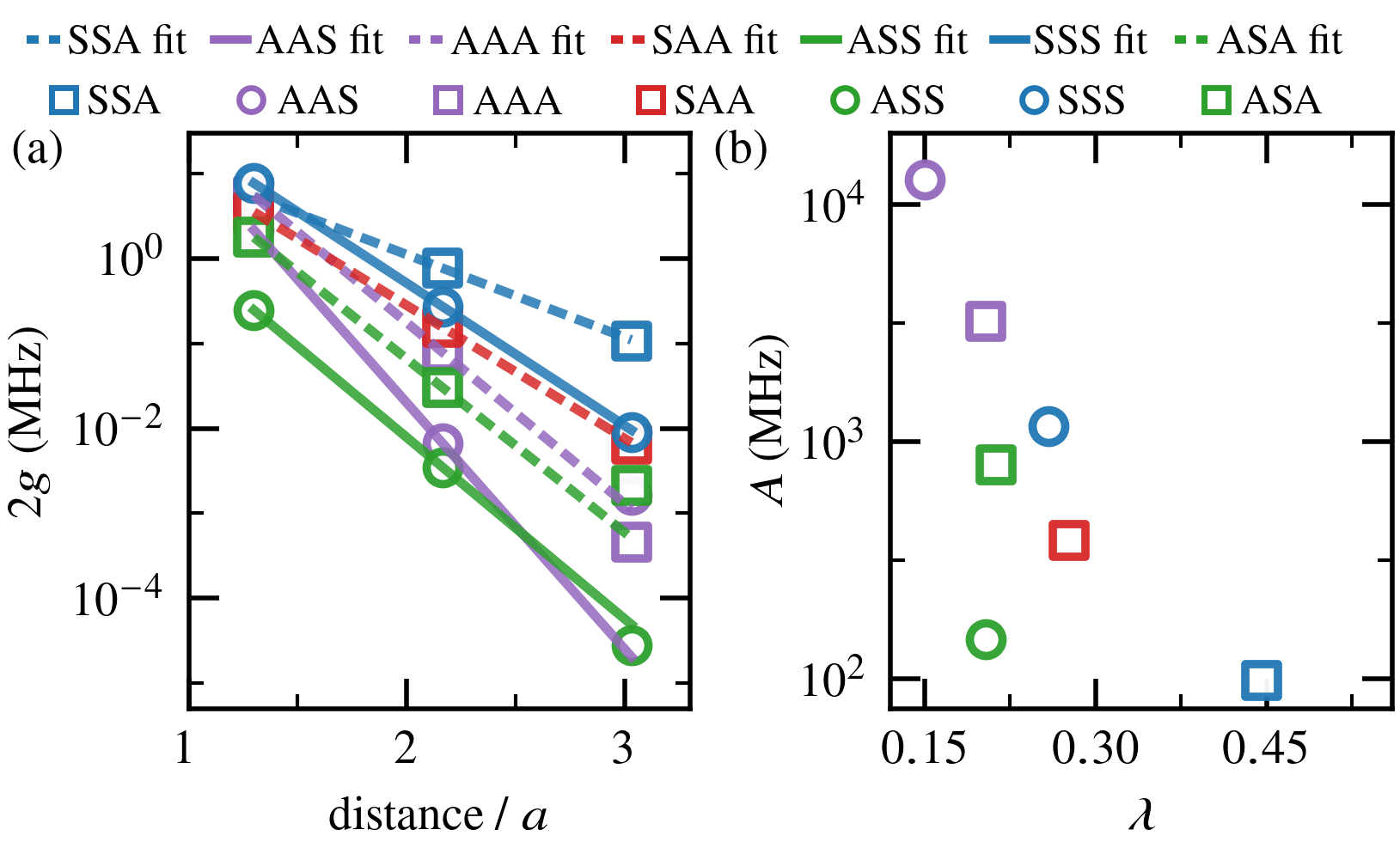}
	\caption{
    (a) Coupling strength $2g$ as a function of the normalized cavity--waveguide separation $d/a$ for the L0 cavity in configurations M1--M3. Symbols denote the values obtained for the different cavity modes, while the solid and dashed lines show exponential fits of the form $2g(d)=A\exp[-d/(a\lambda)]$. The vertical axis is logarithmic. (b) Fitted amplitude $A$ as a function of the decay length $\lambda$ for the same cavity modes. The vertical axis is logarithmic.}
	\label{fig:fig6}
\end{figure}
Finally, we investigate the dependence of the coupling strength~$2g$ on the distance between the cavity and the waveguide for the M1--M3 separation configurations, focusing on the selected modes and anticrossings that displayed the highest coupling strengths. As shown on the logarithmic vertical scale in Fig.~\ref{fig:fig6}(a), the observed dependence is well described by single-exponential fits of the form~$2g(d) = A \exp(-d/(a\lambda))$, shown by the solid and dashed lines in Fig.~\ref{fig:fig6}, where $\lambda$ is the decay length expressed in units of the lattice constant $a$. The fitted values of~$\lambda$ are reported in Table~\ref{tab:fits}.

\begin{table}[tb]
\begin{ruledtabular}
\begin{tabular}{lcc}
\textrm{Mode} & A (MHz) & $\lambda$ \\
\colrule
SSA     & $98.82$          & $0.4458$  \\
AAS     & $1.27\times 10^{4}$ & $0.1497$  \\
AAA     & $3.220\times 10^{3}$ & $0.2038$  \\
SAA     & $383.5$          & $0.2770$  \\
ASS     & $147.7$          & $0.2032$  \\
SSS     & $1.165\times 10^{3}$ & $0.2588$  \\
ASA     & $801.1$          & $0.2126$  \\
\end{tabular}
\end{ruledtabular}
\caption{\label{tab:fits}%
Fitting parameters of $2g(d) = A e^{-d/(a\lambda)}$ for each mode. The parameter $\lambda$ represents the decay length expressed in units of the lattice constant $a$.}
\end{table}

To further quantify the relationship between the fit parameters, Fig.~\ref{fig:fig6}(b) shows the extracted amplitude \( A \) as a function of the decay length \( \lambda \). The extracted parameters $A$ and $\lambda$ show an overall inverse tendency. As evident from Fig.~\ref{fig:fig6}(b), modes with large short-range coupling amplitudes $A$, such as AAS and AAA, exhibit short decay lengths $\lambda$, whereas SSA exhibits a much smaller $A$ and a longer decay length.
In particular, the AAS mode exhibits the largest amplitude ($A \approx 1.27\times10^{4}$~MHz) and the shortest decay length ($\lambda \approx 0.15$), whereas the SSA mode shows the smallest amplitude ($A \approx 99$~MHz) and the longest decay length ($\lambda \approx 0.45$). The remaining modes exhibit intermediate values of both parameters. Given that the exponential fits are based on the three investigated cavity--waveguide separations, these values should primarily be regarded as a compact characterization of the observed distance dependence rather than as a universal classification of the modes.

\subsection{L1 cavity}
\begin{figure}[tb]
	\includegraphics[width=\columnwidth]{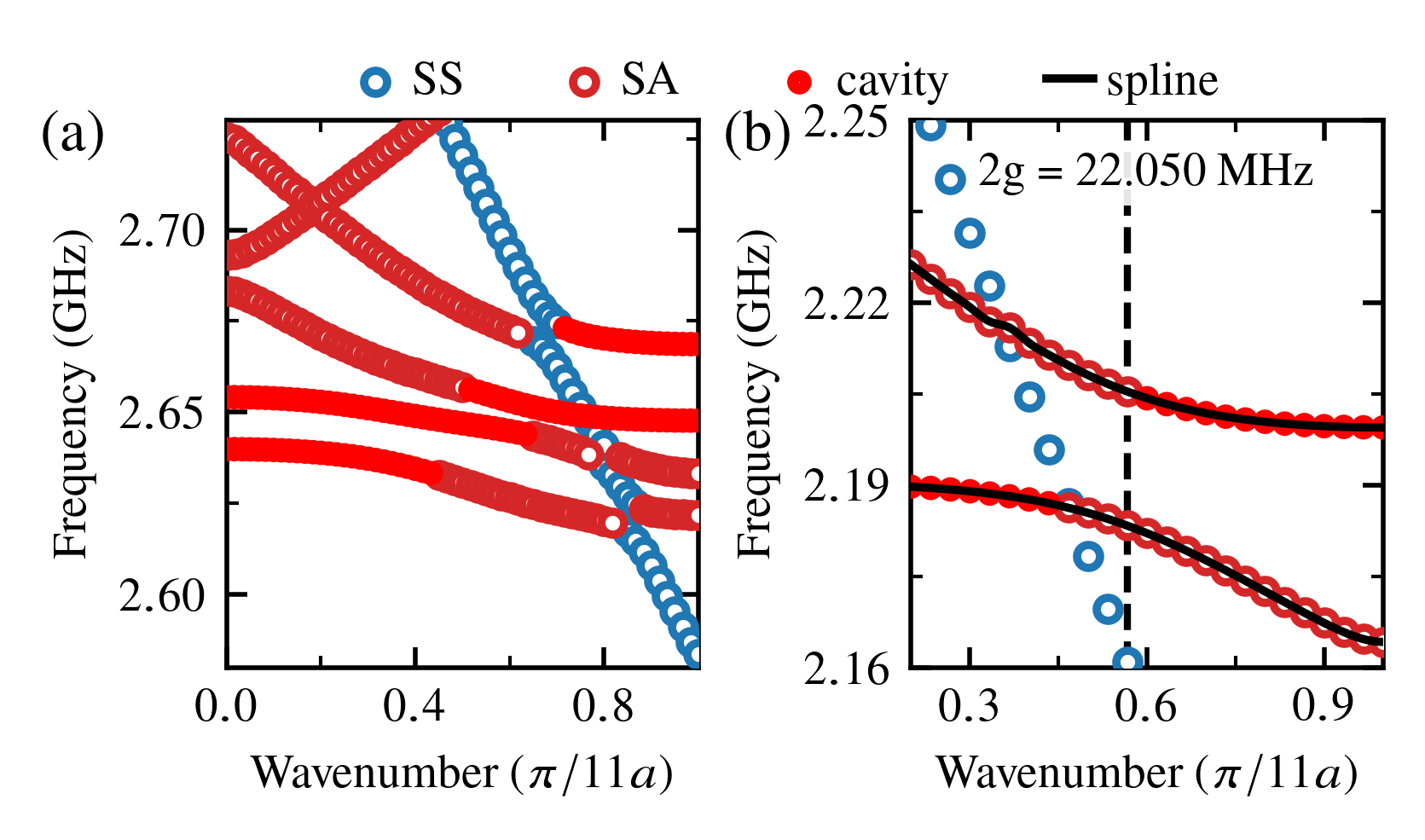}
	\caption{Dispersion relations for the coupled L1-cavity--W1m-waveguide system. (a) Representative case showing multiple overlapping anticrossings in which several cavity (here SSA and SAA) and waveguide modes hybridize simultaneously, resulting in a strongly perturbed spectrum. (b) Exceptionally clean avoided crossing involving the SSS cavity mode, for which a reliable coupling strength can be extracted.}\label{fig:fig7}
\end{figure}
Finally, we extend the analysis to the L1 cavity, which supports nearly twice as many localized modes as the L0 design and therefore exhibits a qualitatively different coupling regime due to its increased modal density. This increased modal density leads to a significantly more perturbed and complex interaction landscape. In most cases, more than two modes hybridize simultaneously [see Fig.~\ref{fig:fig7}(a)], which complicates the identification of isolated anticrossings. Figure~\ref{fig:fig7}(a) illustrates a representative situation where two closely spaced cavity modes of the same $\sigma_z$ symmetry, the SSA mode at $\sim 2.64$~GHz and the SAA mode at $\sim 2.66$~GHz, simultaneously hybridize with the SS and SA waveguide branches within a narrow frequency range. As a result, clean avoided crossings are rarely resolved, which generally precludes reliable extraction of individual coupling strengths $g$ for the L1 cavity. Only in rare cases can individual anticrossings be clearly resolved. Figure~\ref{fig:fig7}(b) presents one such case for the first mode of the L1 cavity, which has SSS symmetry. Although three modes interact in this region (SS and SA waveguide branches together with the cavity mode), the SS branch couples only very weakly. This weak interaction does not significantly perturb the much stronger coupling between the SA waveguide branch and the cavity mode. Consequently, a well-isolated anticrossing is observed, allowing reliable extraction of the coupling strength $2g \approx 22$\,MHz. This value is substantially larger than the coupling extracted for the ASA mode of the smaller L0 cavity.

\section{Experimental realization}\label{sec:results_experiment}
The coupling parameters extracted from the eigenfrequency simulations provide a detailed quantitative picture, but they rely on idealized, defect-free geometries. An experimental realization is therefore important to assess the practical feasibility of these coupled cavity--waveguide structures in a real device. Here, we exploit the optomechanical coupling between integrated InGaAs self-assembled QDs embedded in the center of the membrane and elastic waves to probe the mechanical modes of a representative cavity structure. This approach provides a well-established experimental method for characterizing nanomechanical systems~\cite{nysten2017multi,vogele2020quantum}. In brief, mechanical vibrations modulate the local strain field and thereby shift the QD transition energy via deformation potential coupling ~\cite{gell2008modulation, metcalfe2010resolved, schulein2015fourier, nysten2017multi, weiss2018interfacing, wigger2021resonance, rosinski2026couplingquantumdotselastic}. In time-integrated photoluminescence (PL) spectroscopy, this spectral modulation leads to a characteristic broadening of the QD emission line, from which the modulation amplitude $\Delta E$ can be extracted~\cite{nysten2017multi, weiss2018multiharmonic, vogele2020quantum}. Further details on the device fabrication and the optical detection setup are provided in Appendix~\ref{sec:appendix_experiment}.

For the experimental realization, we fabricate a snowflake-type PhC in a $220\,\mathrm{nm}$ thick GaAs membrane with lattice constant $a=4\,\mu\mathrm{m}$, hole radius $r=1800\,\mathrm{nm}$, and hole width $w=600\,\mathrm{nm}$. These parameters yield phononic band gaps of approximately $361$--$605$~MHz and $338$--$406$~MHz for modes symmetric and antisymmetric with respect to $\sigma_z$, respectively, resulting in a common band gap of $361$--$406$~MHz. Within this PhC, a W1m waveguide is formed using the offset $\Delta=\sqrt{3}\,a/4$, resulting in a core width $b=\sqrt{3}\,a/2-w$. The waveguide is coupled to an L2 cavity, which was chosen for its experimental robustness, as its modes are less sensitive to fabrication imperfections. The M2 cavity--waveguide separation was chosen to limit excessive hybridization and facilitate the identification of individual cavity resonances. Compared with the structures considered in Sec.~\ref{sec:results}, which operate near 3~GHz, the larger lattice constant shifts the phononic band gap and cavity resonances to the sub-GHz range, where they are readily accessible via surface acoustic wave (SAW) excitation.

Excitation of the cavity modes is achieved using Rayleigh SAWs generated by chirped interdigital transducers (IDTs) on the bulk substrate. The spatially varying IDT periodicity enables SAW excitation over a broad frequency range~\cite{weiss2018multiharmonic}. Upon reaching the suspended membrane, the waves convert into plate modes that propagate through the waveguide and couple to the cavity. This conversion further restricts the set of relevant waveguide modes, as the resulting plate modes selectively excite only a subset of the available waveguide modes. In particular, the excitation predominantly addresses modes whose second symmetry label is S, i.e., modes that are symmetric with respect to the $\sigma_y$ plane (SS and AS)~\cite{korovin2019conversion}. The SA and AA branches are not efficiently excited in this configuration and are therefore not expected to contribute significantly to the experimental response. Further details of the radio-frequency excitation are provided in Appendix~\ref{sec:appendix_experiment}.

\begin{figure*}[tb]
	\includegraphics[width=\textwidth]{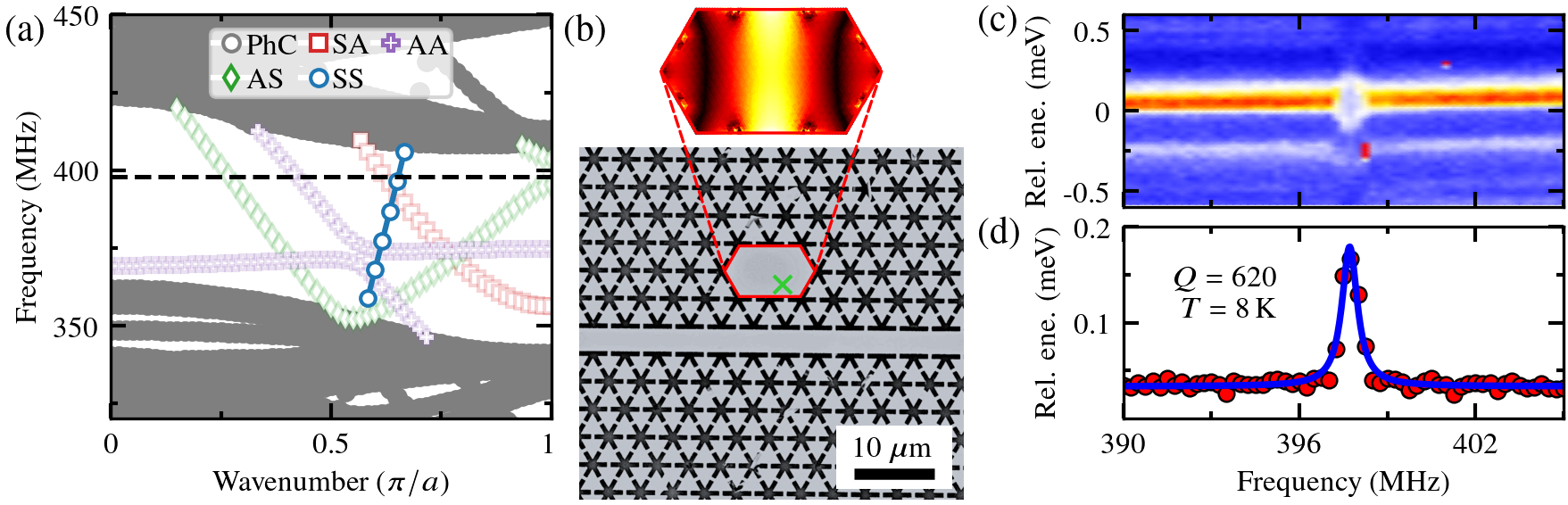}
	\caption{(a) Dispersion relations of the guided modes in the PhC waveguide used in the experiment. Modes that are not efficiently excited in the employed configuration (SA and AA) or are symmetry-forbidden from coupling to the SSS cavity mode (AS) are shown in pale colors. The gray shaded regions denote the PhC bands (PC), and the black dashed line marks the experimentally observed cavity resonance. (b) SEM image of the cavity--waveguide structure, with the measurement position marked by a green cross. The inset shows the simulated volumetric strain of the cavity mode at approximately 395~MHz. (c) False-color plot of the QD's time-integrated emission line measured inside the cavity as a function of the SAW frequency. (d) Extracted modulation amplitude $\Delta E$ as a function of the SAW frequency, }\label{fig:fig8}
\end{figure*}

The experimental response is further narrowed by the selectivity of the QD detection mechanism. The QDs are located in the mid-plane of the membrane ($z=0$), where AS modes exhibit vanishing volumetric strain~\cite{rosinski2026couplingquantumdotselastic}. Their coupling to the QDs can therefore occur only through off-diagonal strain components, for which the resulting modulation of the QD transition energy is quadratic and thus gives rise to a frequency-doubled optical response~\cite{rosinski2026couplingquantumdotselastic}. Since such a response is not observed experimentally, the contribution of AS modes is strongly suppressed. In contrast, SS modes exhibit nonzero volumetric strain in the mid-plane~\cite{rosinski2026couplingquantumdotselastic}, enabling efficient deformation-potential coupling to the QDs.

Figure~\ref{fig:fig8}(b) shows a scanning electron microscope (SEM) image of the fabricated cavity--waveguide structure, with the measurement position inside the cavity marked by a green cross. Figure~\ref{fig:fig8}(c) shows the QD's time-integrated emission line as a function of the SAW frequency in false-color representation, where a pronounced resonance is observed around 398~MHz. The modulation amplitude $\Delta E$ is extracted from these data and plotted as a function of the SAW frequency in Fig.~\ref{fig:fig8}(d), where the resonance is fitted with a Lorentzian function. Further details of the measurement and data-analysis procedure are provided in Appendix~\ref{sec:appendix_experiment}. The resulting resonance curve yields a mechanical quality factor of $Q=620$, consistent with literature values for similar GaAs phononic structures~\cite{hatanaka2020real}. Additional measurements performed directly on the waveguide and the surrounding PhC are presented in Appendix~\ref{sec:additionalMeasurement}. These control measurements confirm the suppression of mechanical-wave propagation through the surrounding PhC within the band gap, while a finite response persists in the waveguide, supporting the interpretation that the cavity is excited through guided waveguide modes.

\begin{figure}[tb]
	\includegraphics[width=\columnwidth]{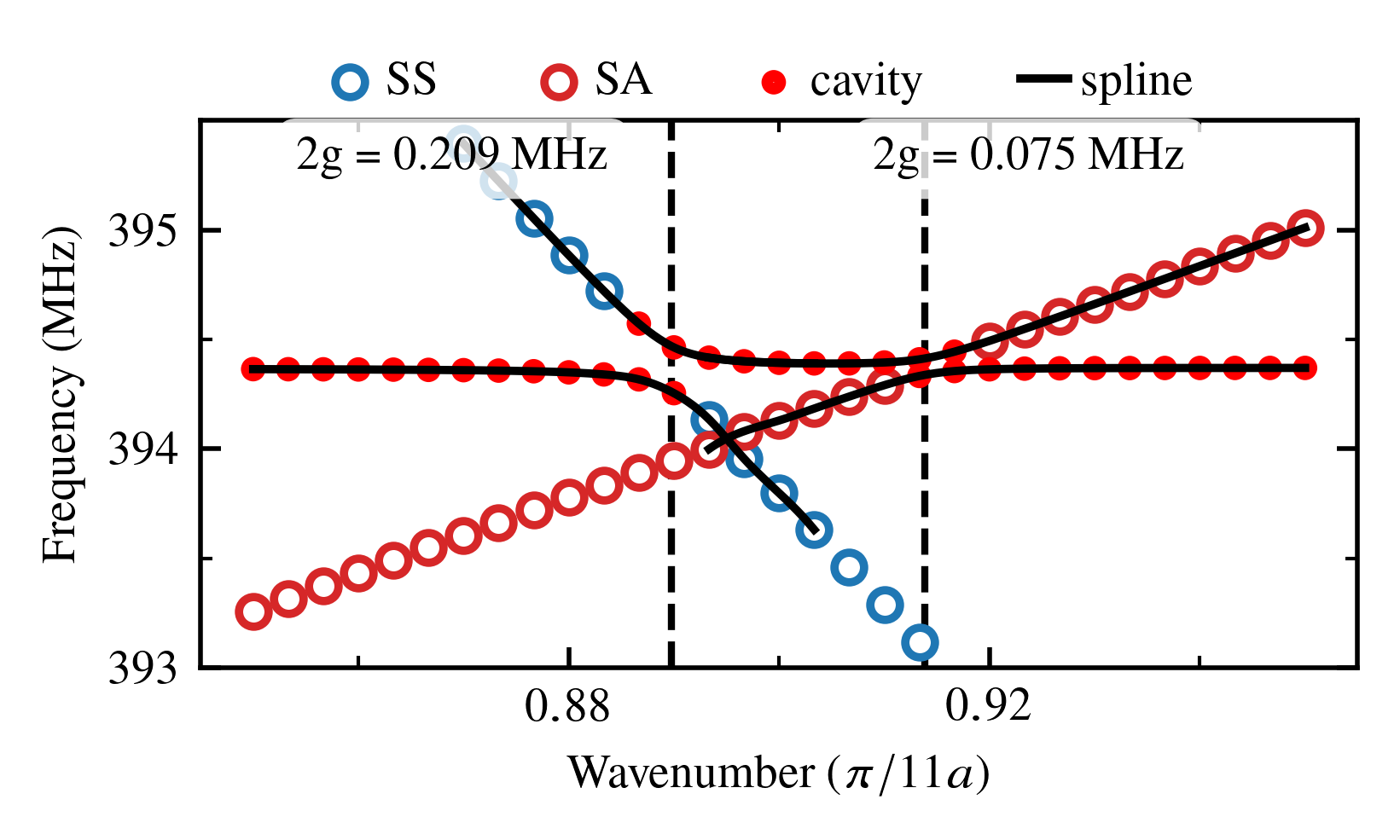}
	\caption{FEM dispersion diagram illustrating the avoided crossings between the SSS cavity mode and the SS and SA waveguide modes. Red filled dots denote modes identified as cavity-like, while blue and red open circles correspond to the SS and SA waveguide modes, respectively. Black solid lines are cubic-spline fits used to extract the minimum frequency splitting, and black dashed vertical lines mark the Bloch wavenumbers at which the avoided crossings occur. }
	\label{fig:fig9}
\end{figure}

To interpret the experimentally observed resonance, we compare the measurements with numerical simulations of the cavity and waveguide modes. Figure~\ref{fig:fig8}(a) shows the calculated dispersion of the PhC waveguide together with the symmetry classification of the relevant guided modes. Among the cavity modes, the SSS mode is the only one located in the vicinity of the experimentally observed resonance at 398~MHz. Figure~\ref{fig:fig9} therefore presents the FEM dispersion diagram in the vicinity of this mode, revealing its interaction with the SS and SA waveguide modes. The resulting avoided crossings occur at Bloch wavenumbers $k = 0.88976\,(\pi/11a)$ and $k = 0.91389\,(\pi/11a)$, with minimum frequency splittings of $2g = 0.209$~MHz and $2g = 0.075$~MHz, respectively. Expressed in terms of the primitive unit cell of length $a$, these interactions take place at approximately $k \approx 0.64\text{--}0.65\,\pi/a$. The SSS--SS anticrossing occurs at approximately 395~MHz, in good agreement with the experimentally observed resonance at 398~MHz, supporting the identification of the experimental resonance with the simulated SSS cavity mode. The simulated volumetric strain of this cavity mode is shown in the inset of Fig.~\ref{fig:fig8}(b) and exhibits a comparatively large amplitude at the measurement position, consistent with efficient deformation-potential coupling to the embedded QD. Consequently, although both SS and SA waveguide branches couple to the SSS cavity mode in the FEM calculations, the experimentally observed response is expected to be dominated by the SS branch.

\section{Conclusion}\label{sec:conclusion}
We have investigated the coupling between localized mechanical modes of snowflake-type PhC cavities and propagating modes of a W1m PhC waveguide using FEM eigenfrequency simulations complemented by experimental measurements in a GaAs membrane. The numerical analysis shows that the coupling strength is strongly correlated with the spatial overlap integral between the displacement fields of the corresponding isolated cavity and waveguide modes. After accounting for the effective-mass dependence of the normalized cavity displacement amplitude, this overlap provides a leading-order description of the interaction strength across a wide range of mode symmetries and configurations. Deviations from this leading-order behavior are associated with Bloch-phase effects, waveguide dispersion associated with variations in group velocity, and intrinsic hybridization of waveguide modes.

Analysis of the distance dependence reveals an interplay between coupling amplitude and spatial decay length. Extension of the analysis to the larger L1 cavity shows that its increased modal density can lead to simultaneous hybridization of several cavity and waveguide modes over overlapping spectral ranges. Such multimode hybridization generally prevents the reliable extraction of individual coupling strengths from isolated avoided crossings. Our results therefore highlight two complementary strategies for obtaining well-defined cavity--waveguide coupling: reducing the number of available waveguide modes in the relevant frequency range, ideally approaching a single-mode regime, or reducing the interaction strength by increasing the cavity--waveguide separation.

Finally, experimental measurements on a GaAs PhC membrane provide proof-of-principle evidence consistent with cavity--waveguide coupling through a pronounced resonance in the spectral broadening of the PL emission from an embedded QD, which serves as a local probe of the mechanical field. The observed resonance at approximately 398~MHz is in good agreement with the simulated SSS--SS avoided crossing near 395~MHz. These results demonstrate the practical feasibility of coupling localized and propagating mechanical modes in PhC structures.

These insights provide practical guidelines for engineering cavity--waveguide interactions by controlling mode overlap, cavity--waveguide separation, and the modal structure of the coupled components. Such control is important for achieving selective transfer of mechanical excitations between localized cavity modes and propagating waveguide modes, providing a basis for the design of more complex on-chip phononic networks.

\begin{acknowledgements}
This project was supported by the German Federal Ministry of Education and Research via the Research Group Linkage Program of the Alexander von Humboldt Foundation. J.~R. and P.~M. acknowledge funding from the Narodowe Centrum Nauki (NCN, Polish National Science Centre), grant no. 2023/50/A/ST3/00511. H.~J.~K. and B.~M. acknowledge support by the Deutsche Forschungsgemeinschaft (DFG, German Research Foundation), projects 465136867, 505596454, 563184308.
\end{acknowledgements}

\appendix

\section{Band folding}\label{sec:appendix_folding}

We consider a one-dimensional periodic waveguide with primitive lattice constant $a$. Bloch modes of the primitive cell are characterized by a wavenumber
\[
k \in \left[-\frac{\pi}{a},\,\frac{\pi}{a}\right]
\]
within the first BZ. To describe the same waveguide using an enlarged computational domain, we construct a supercell of length $L = N a$, where $N$ is a positive integer, and impose Floquet--Bloch boundary conditions with respect to this enlarged periodicity. The corresponding BZ is reduced to
\[
k_L \in \left[-\frac{\pi}{N a},\,\frac{\pi}{N a}\right].
\]

The change in periodicity leads to band folding, since wavenumbers differing by integer multiples of the supercell reciprocal lattice vector
\[
G_L = \frac{2\pi}{N a}
\]
are physically equivalent. Consequently, the wavenumber $k$ of a primitive cell Bloch mode reappears in the supercell at a wavenumber
\[
k_L = k - n G_L,
\]
where the integer $n$ is chosen such that
\[
k_L \in \left[-\frac{\pi}{N a},\,\frac{\pi}{N a}\right].
\]
This reduction uniquely determines the location of the folded Bloch mode within the supercell BZ. While the corresponding folded mode retains the eigenfrequency $\omega(k)$ of the original primitive-cell Bloch mode, the folding procedure introduces an important complication: the supercell eigenproblem generally exhibits degeneracies. For a supercell containing $N$ primitive cells, $N$ distinct wavenumbers from the primitive-cell BZ fold onto the same wavenumber $k_L$ in the reduced supercell BZ. As a result, numerical solvers may return arbitrary linear combinations of these degenerate Bloch modes, whose field profiles may differ from the primitive-cell mode despite sharing the same eigenfrequency.

To consistently compare folded modes obtained for different supercell sizes, we express the folded wavenumber in dimensionless units of the corresponding supercell Brillouin-zone edge
\[
\tilde{k}_L = \frac{k_L}{\pi/(N a)} \in [-1,1].
\]
This representation is used throughout this work to identify and track Bloch modes across different supercell sizes.

\section{Effective masses and zero-point motion of the L0 cavity modes}\label{sec:appendix_cavityOpto}
The effective mass $m_{\rm{eff}}$ determines the displacement amplitude of a mechanical mode for a given mechanical energy and is therefore relevant to the normalization of the cavity displacement fields used in the overlap analysis presented in Sec.~\ref{sec:results}. Defining the mode amplitude by the maximum displacement magnitude, the effective mass is given by~\cite{eichenfield2009modeling,aspelmeyer2013cavity}
\begin{equation}
m_{\mathrm{eff}}=
\frac{\int \rho |\mathbf{u}(\mathbf{r})|^2\,dV}
{\max[|\mathbf{u}(\mathbf{r})|^2]}.
\end{equation}
The effective mass quantifies the spatial extent of the displacement field relative to its maximum amplitude. As can be seen in Appendix~\ref{sec:appendix_fields}, the modes with large effective masses (AAA, SSS, and ASA) exhibit broadly distributed displacement fields, whereas the modes with small effective masses show pronounced displacement maxima localized within a relatively small volume.

For completeness, we also report the corresponding zero-point motion amplitudes
\begin{equation}
x_{\mathrm{zpf}}=\sqrt{\frac{\hbar}{2m_{\mathrm{eff}}\omega}}.
\end{equation}

For the L0 cavity modes considered here, the calculated effective masses and corresponding zero-point motion amplitudes are summarized in Table~\ref{tab:effective_mass}. The effective masses range from approximately $17$ to $169$~fg, with zero-point motion amplitudes reaching approximately $13$~fm.

\begin{table}[tb]
\begin{ruledtabular}
\begin{tabular}{cccc}
Freq. (GHz) & symmetry & $m_{\text{eff}}$ (fg) & $x_{\text{zpf}}$ (fm) \\
\colrule
2.41  & SSA & 19.194  & 13.466 \\
2.71  & AAS & 18.825  & 12.824 \\
2.75  & AAA & 118.07  & 5.0771 \\
3.06  & SAA & 16.826  & 12.778 \\
3.14  & ASS & 25.358  & 10.268 \\
3.22  & SSS & 102.94  & 5.0262 \\
3.29  & ASA & 169.10  & 3.8777 \\
\end{tabular}
\end{ruledtabular}
\caption{Effective masses $m_{\mathrm{eff}}$ and zero-point motion amplitudes $x_{\mathrm{zpf}}$ of the localized L0 cavity modes.}
\label{tab:effective_mass}
\end{table}

\section{Extraction of coupling strength}\label{sec:appendix_coupling_extraction}
The coupling strength between cavity and waveguide modes is extracted from avoided crossings in the dispersion relation of the coupled system. For each anticrossing, the eigenfrequencies are grouped into continuous branches using a mode-tracking algorithm and interpolated using cubic splines with a smoothing parameter $s=0$. The coupling strength is then determined as the minimum frequency separation between the two interpolated branches,
\begin{equation}
2g = \min_k |f_1(k) - f_2(k)|,
\end{equation}
where $f_1(k)$ and $f_2(k)$ denote the interpolated dispersion relations. The minimum is evaluated numerically on a dense grid of $k$-points and refined within a narrow interval around the initial minimum.

\section{Field profiles}\label{sec:appendix_fields}
\begin{figure*}[p]
	\includegraphics[width=\textwidth]{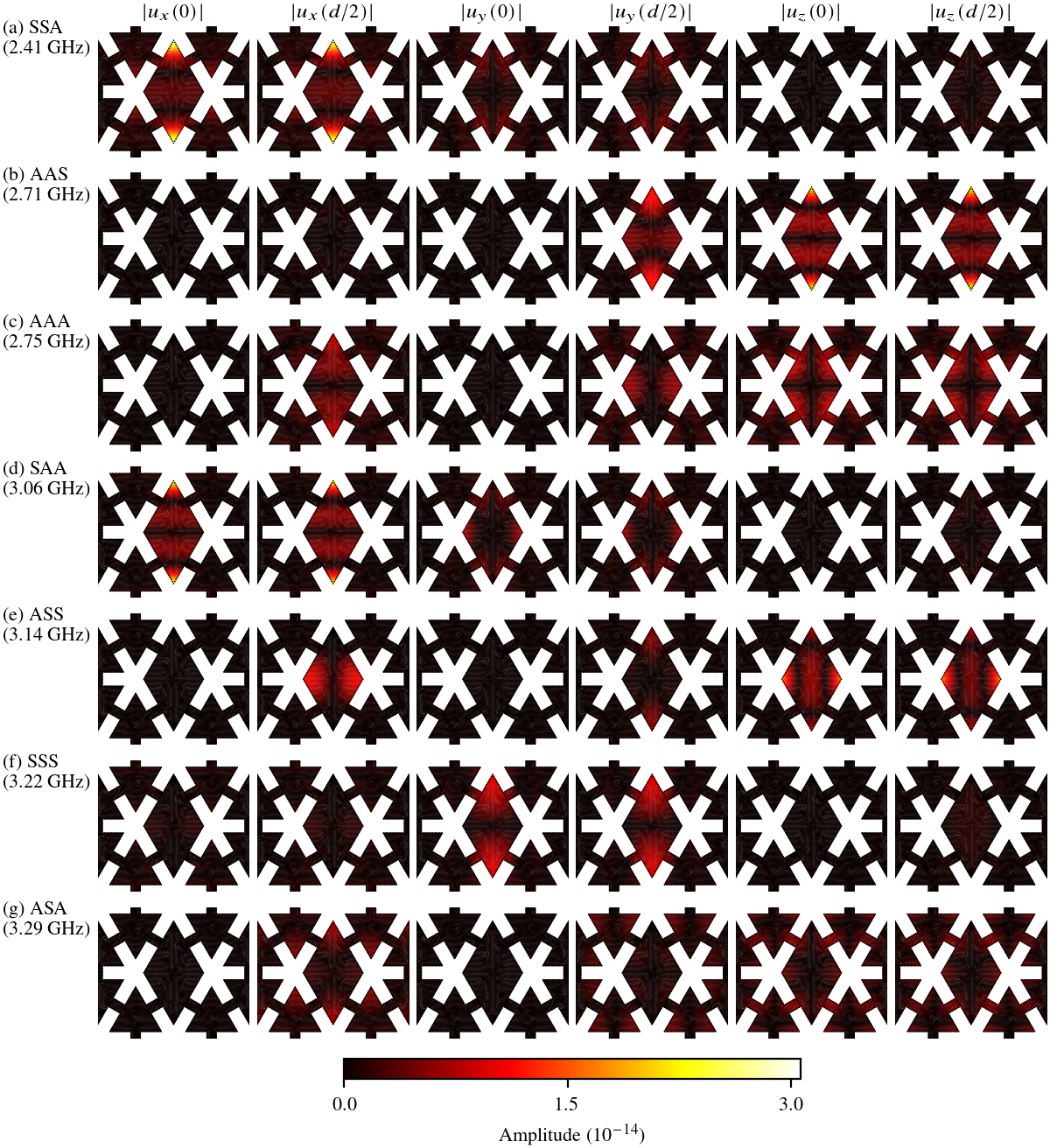}
    \caption{Spatial distribution of the amplitudes of the displacement-field components for the first seven modes of the L0 cavity, ordered by increasing frequency from (a) to (g). The three pairs of columns show the $u_x$, $u_y$, and $u_z$ components, respectively. Within each pair, the left and right columns correspond to horizontal cross-sections at $z=0$ and $z=h/2$, respectively. The color scale represents the normalized displacement amplitude.}
	\label{fig:fig10}
\end{figure*}

\begin{figure*}[p]
	\includegraphics[width=\textwidth]{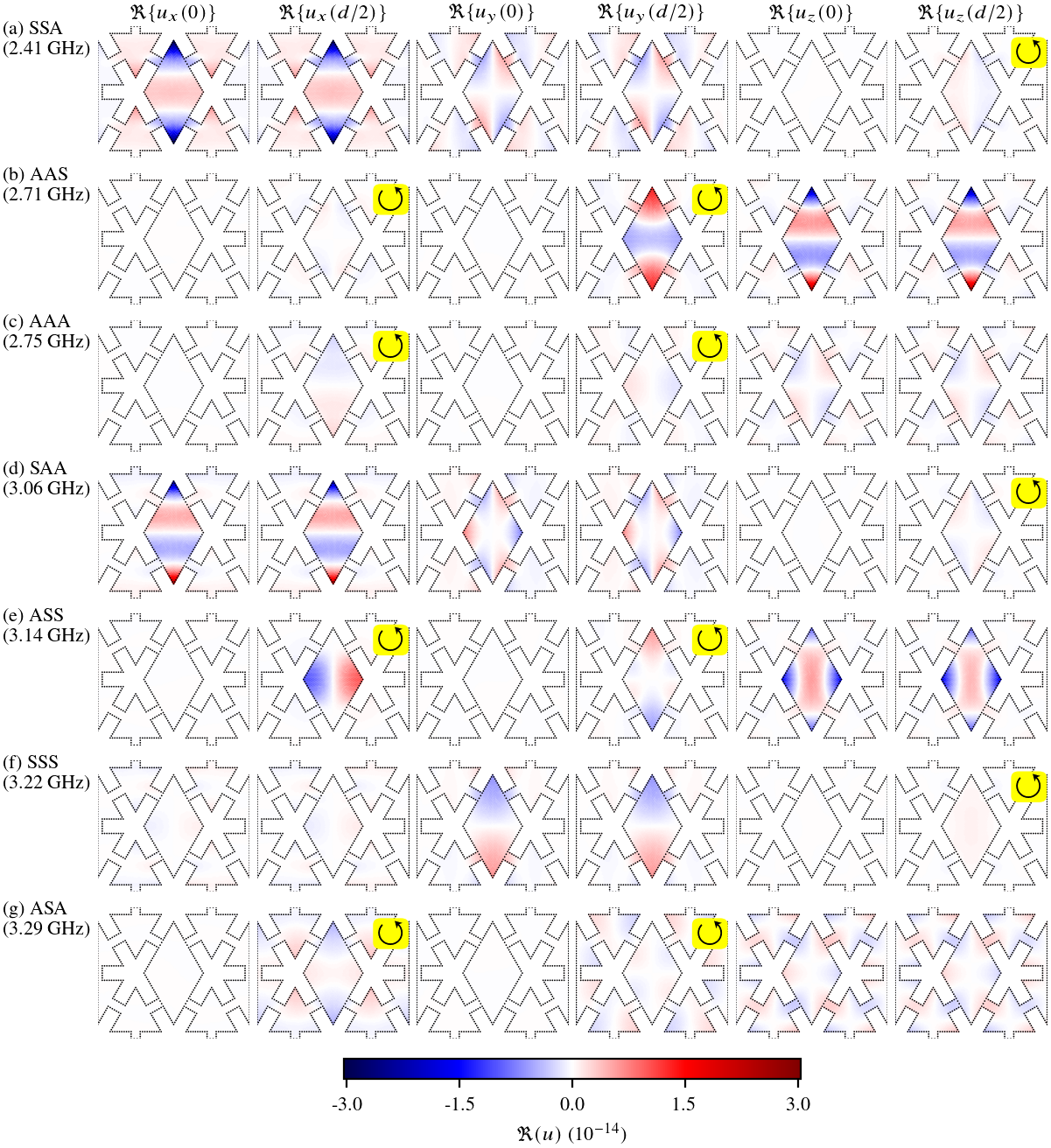}
    \caption{Spatial distribution of the real parts of the displacement-field components for the first seven modes of the L0 cavity, ordered by increasing frequency from (a) to (g). The three pairs of columns show the $u_x$, $u_y$, and $u_z$ components, respectively. Within each pair, the left and right columns correspond to horizontal cross-sections at $z=0$ and $z=h/2$, respectively. The color scale represents the normalized real part of the displacement field. The yellow circular arrow denotes fields that change sign upon translation from $z=h/2$ to $z=-h/2$.}
	\label{fig:fig11}
\end{figure*}

This Appendix presents the displacement-field profiles of the seven L0 cavity modes considered in the main text. Figures~\ref{fig:fig10} and \ref{fig:fig11} show the three displacement components $u_x$, $u_y$, and $u_z$. The displacement fields are normalized to the same mechanical energy, enabling a direct comparison of the displacement amplitudes across different modes. For each component, two adjacent columns correspond to horizontal cross-sections at $z=0$ and $z=h/2$, respectively, allowing both the in-plane spatial distribution and its variation across the membrane thickness to be examined. The first, second, and third pairs of columns correspond to $u_x$, $u_y$, and $u_z$, respectively. Figure~\ref{fig:fig10} shows the displacement amplitudes, whereas Fig.~\ref{fig:fig11} shows the real parts of the displacement fields, additionally revealing their relative signs and symmetry properties.

\section{Experimental details}\label{sec:appendix_experiment}

\subsection*{Device fabrication}
The studied devices were fabricated on a MBE-grown heterostructure following the procedure described in our previous work \cite{vogele2020quantum}: IDTs are fabricated on the bulk GaAs substrate using positive tone electron-beam lithography and a lift-off process. After the fabrication of the IDTs, PnCs were patterned in another electron-beam lithography step in the propagation path of the SAWs. The PnC pattern was transferred into the heterostructure by inductively coupled plasma reactive ion etching (ICP-RIE) using a chlorine based etch chemistry. Finally, PnC membranes are suspended by selective etching of a AlAs sacrificial layer using hydrofluoric acid. 

\subsection*{SAW excitation}
SAWs were generated using a chirped split-52 IDT \cite{weiss2018multiharmonic,weiss2021optomechanical}. The IDT was designed to operate over a frequency range of approximately 300--550\,MHz, covering the phononic band-gap region of interest. The radio-frequency (RF) signal was supplied by a standard RF signal generator (Stanford Research Systems SG382) and amplified by a low-noise amplifier (Mini-Circuits ZHL-2010+). For the PhC and waveguide measurements, an RF power of 25\,dBm was applied, while cavity measurements were performed at 27\,dBm. Frequency scans were performed in discrete steps of 0.25\,MHz to resolve narrow mechanical resonances. To reduce heating of the sample and minimize temperature drift effects, SAWs were generated in short pulses with a typical pulse length of 2\,$\mu$s and a repetition period of 20\,$\mu$s. The such generated SAW couple into the suspended PnC membrane~\cite{fuhrmann2011dynamic,kapfinger2015dynamic, balram2016coherent, vogele2020quantum}.

\begin{figure*}[tb]
	\includegraphics[width=\textwidth]{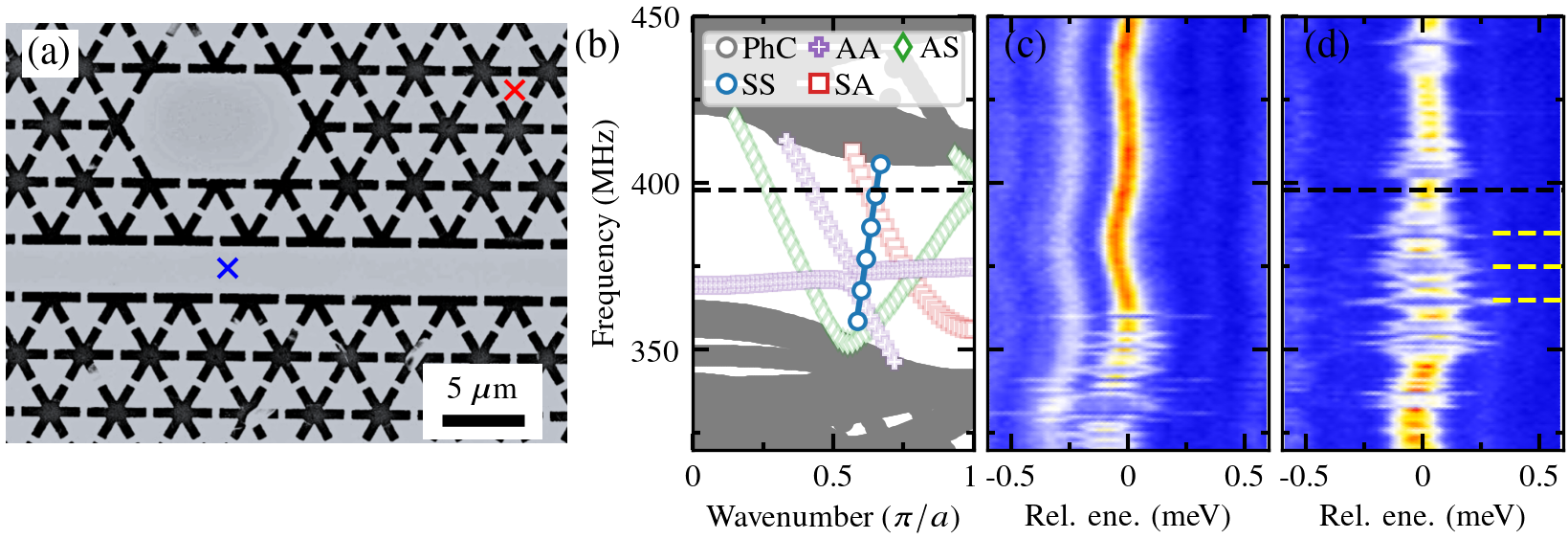}
	\caption{(a) SEM image with corresponding measurement locations on the waveguide (blue cross) and the PhC (red cross). (b) Computed dispersion relations of the guided modes in the PhC waveguide used in the experiment. Modes that are either symmetry-forbidden under the employed excitation (SA and AA) or symmetry-forbidden from coupling to the SSS cavity mode (AS) are shown in pale colors. The gray shaded regions denote PhC bands (PC). (c,d) False-color plots of the QD's time-integrated emission line measured on the PhC and the waveguide, respectively. In panels (b) and (d), the black dashed line marks the frequency of the cavity mode determined experimentally from the measurements shown in the main text. The three yellow dashed lines in panel (d) correspond to frequencies of 365, 375, and 385~MHz.}\label{fig:fig12}
\end{figure*}

\subsection*{Optical detection}
The mechanical response of the cavity was probed optically using self-assembled InGaAs QDs embedded in the  center of the membrane. All measurements were performed in a Helium flow cryostat at a bath temperature of approximately 8\,K. The excitation and collection were performed in a confocal micro-PL setup. The sample was excited using a pulsed diode laser (Pico Quant LDH-D-C-660) emitting 90\,ps long pulses at a wavelength of 660\,nm  at a repetition rate of 80\,MHz. The emitted luminescence of the QD was spectrally dispersed in a 0.5\,m grating spectrometer and detected by a cooled Silicon CCD detector. 

\subsection*{Measurement procedure}
For each measurement, an individual QD was selected and its emission line was monitored while sweeping the RF excitation frequency of the IDT. At each frequency step, a PL spectrum was recorded and the  modulation amplitude  $\Delta E$ was extracted by fitting the spectral line with a Lorentzian profile. Mechanical cavity resonances were identified as sharp maxima in the frequency-dependent  modulation amplitude $\Delta E(f)$. The mechanical quality factor was determined by fitting the resonance peaks with a Lorentzian function in the frequency domain and extracting the resonance frequency $f_{\mathrm{res}}$ and the full width at half maximum $\Delta f$, yielding $Q = f_{\mathrm{res}}/\Delta f$.

\section{Measurements on a PhC and a PhC Waveguide}\label{sec:additionalMeasurement}

To complement the cavity measurements presented in the main text, we performed additional measurements directly on the PhC and on the PhC waveguide. The measurement positions are illustrated in the SEM image in Fig.~\ref{fig:fig12}(a). As in the cavity measurements presented in the main text, the QDs used as local strain sensors are located in the membrane mid-plane at $z=0$. These measurements provide a reference for identifying the response associated with propagating PhC and waveguide modes and distinguishing it from cavity-specific resonances. For reference, Fig.~\ref{fig:fig12}(b) shows the same calculated  dispersion of the PhC waveguide as in the main text.

Figure~\ref{fig:fig12}(c) presents the response measured directly on the PhC. Below the phononic band gap, a clear modulation of the QD emission is observed due to coupling to the extended PhC modes. Within the phononic band gap, this response disappears, consistent with the absence of propagating PhC modes in this frequency range. Above the band gap, coupling to the PhC modes is again  expected; however, no modulation is observed as fabrication imperfections limit efficient SAW excitation in this frequency range.

Figure~\ref{fig:fig12}(d) shows the response measured on the PhC waveguide. In the frequency regions corresponding to the PhC bands, no pronounced waveguide-related QD modulations are observed. In contrast, several distinct features appear within the phononic band gap, where the SS guided mode exists. As shown by the dispersion relation in Fig.~\ref{fig:fig12}(b), this mode extends across the band gap and therefore provides guided states at different frequencies and Bloch wavevectors. The linewidth-broadening features observed within the band gap occur in the frequency range supported by the SS guided mode and are therefore attributed to the waveguide response.
\FloatBarrier

\bibliography{paper}

\end{document}